\documentclass[11pt,a4paper]{article}
\pdfoutput=1
\usepackage{jcappub}

\title{Common Origin of Neutrino Mass, Dark Matter and Dirac Leptogenesis}

\author[a]{Debasish Borah, \note{Corresponding author}}
\affiliation[a,1]{Department of Physics, Indian Institute of Technology Guwahati, Assam-781039, India}
\emailAdd{dborah@iitg.ernet.in}
\author[b]{Arnab Dasgupta}
\affiliation[b]{Institute of Physics, HBNI, Sachivalaya Marg, Bhubaneshwar-751005, India}
\emailAdd{arnab.d@iopb.res.in}

\abstract{
We study the possibility of generating tiny Dirac neutrino masses at one loop level through the \textit{scotogenic} mechanism such that one of the particles going inside the loop can be a stable cold dark matter (DM) candidate. Majorana mass terms of singlet fermions as well as tree level Dirac neutrino masses are prevented by incorporating the presence of additional discrete symmetries in a minimal fashion, which also guarantee the stability of the dark matter candidate. Due to the absence of total lepton number violation, the observed baryon asymmetry of the Universe is generated through the mechanism of Dirac leptogenesis where an equal and opposite amount of leptonic asymmetry is generated in the left and right handed sectors which are prevented from equilibration due to tiny Dirac Yukawa couplings. Dark matter relic abundance is generated through its usual freeze-out at a temperature much below the scale of leptogenesis. We constrain the relevant parameter space from neutrino mass, baryon asymmetry, Planck bound on dark matter relic abundance, and latest LUX bound on spin independent DM-nucleon scattering cross section. We also discuss the charged lepton flavour violation $(\mu \rightarrow e \gamma)$ and electric dipole moment of electron in this model in the light of the latest experimental data and constrain the parameter space of the model.}

\begin{document}
\maketitle
\section{Introduction}
The discovery of the Higgs boson at the Large Hadron Collider (LHC) in 2012 followed by a series of null results from beyond standard model (BSM) searches have repeatedly confirmed validity of the standard model (SM) of particle physics at energies accessible to the LHC. In spite of being the most successful theory of particle physics till date, the SM however fails to explain many observed phenomena which include non-zero neutrino masses and mixing, dark matter and matter-antimatter asymmetry, in particular. Though the Higgs field in the SM is responsible for generating masses of all the charged fermions and weak vector bosons, it has no renormalisable coupling with the neutrinos due to the absence of the right handed neutrino. This keeps the neutrinos massless in the SM and hence no leptonic mixing which is ruled out by the experimental observations of non-zero neutrino masses and large leptonic mixing \cite{PDG, kamland08, T2K, chooz, daya, reno, minos}. The present status of neutrino parameters can be summarised by the recent global fit analysis reported in \cite{schwetz14}  and \cite{valle14}  are shown in table \ref{tab:data1}.
\begin{center}
\begin{table}[htb]
\begin{tabular}{|c|c|c|c|c|}
\hline
Parameters & NH \cite{schwetz14} & IH \cite{schwetz14} & NH \cite{valle14} & IH \cite{valle14} \\
\hline
$ \frac{\Delta m_{21}^2}{10^{-5} \text{eV}^2}$ & $7.02-8.09$ & $7.02-8.09 $ & $7.11-8.18$ & $7.11-8.18 $ \\
$ \frac{|\Delta m_{31}^2|}{10^{-3} \text{eV}^2}$ & $2.317-2.607$ & $2.307-2.590 $ & $2.30-2.65$ & $2.20-2.54 $ \\
$ \sin^2\theta_{12} $ &  $0.270-0.344 $ & $0.270-0.344 $ &  $0.278-0.375 $ & $0.278-0.375 $ \\
$ \sin^2\theta_{23} $ & $0.382-0.643$ &  $0.389-0.644 $ & $0.393-0.643$ &  $0.403-0.640 $ \\
$\sin^2\theta_{13} $ & $0.0186-0.0250$ & $0.0188-0.0251 $ & $0.0190-0.0262$ & $0.0193-0.0265 $ \\
$ \delta $ & $0-2\pi$ & $0-2\pi$ & $0-2\pi$ & $0-2\pi$ \\
\hline
\end{tabular}
\caption{Global fit $3\sigma$ values of neutrino oscillation parameters \cite{schwetz14, valle14}.}
\label{tab:data1}
\end{table}
\end{center}
Although the lightest neutrino mass remains undetermined at neutrino oscillation experiments, cosmology experiments like Planck can put an upper bound on it from the bound they put on the sum of absolute neutrino masses $\sum_i \lvert m_i \rvert < 0.17$ eV \cite{Planck15}. Although the leptonic Dirac CP phase $\delta$ is not yet measured precisely, recently the T2K experiment showed preference for $\delta \approx -\pi/2$ \cite{diracphase}. If neutrinos are Majorana fermions, there exists two Majorana CP phases which remain undetermined at oscillation experiments, but can be probed at neutrinoless double beta decay experiments $(0\nu \beta \beta)$.

The presence of dark matter in the Universe has been known since the observations of galaxy rotation curves made by Fritz Zwicky \cite{zwicky}. Since then, the evidence in favour of dark matter has been increasing with the latest cosmology experiment Planck suggesting around $26\%$ of the present Universe's energy density being made up of dark matter \cite{Planck15}. In terms of density parameter $\Omega$, the Planck bound on dark matter abundance in the present Universe can be written as
\begin{equation}
\Omega_{\text{DM}} h^2 = 0.1187 \pm 0.0017
\label{dm_relic}
\end{equation}
where $h = \text{(Hubble Parameter)}/100$ is a parameter of order unity. Although the astrophysical and cosmological evidences suggesting the presence of dark matter in the Universe have been ever increasing, the particle nature of dark matter is still unknown. It is however confirmed that none of the SM particles can be a dark matter candidate as none of them possess the typical characteristics of dark matter \cite{bertone}. Although the weakly interacting massive particle (WIMP) paradigm is the most well studied dark matter scenario, the experiments like the LHC and the dark matter direct detection experiments \cite{Aprile:2013doa, LUX} have so far been giving null results only. More recently, the LUX experiment has announced \cite{LUX16} a factor of four improvement of its previous exclusion limits, ruling out DM-nucleon spin independent cross section above around $2.2 \times 10^{-46} \; \text{cm}^2$ for DM mass of around 50 GeV.

The matter-antimatter asymmetry in the observed Universe has also been a longstanding puzzle which the SM fails to explain. This observed asymmetry is reported in terms of the baryon to entropy ratio as \cite{Planck15, PDG14}
\begin{equation}
Y_B = \frac{n_B -n_{\bar{B}}}{s}= (1.61-1.83) \times 10^{-10}
\label{barasym}
\end{equation} 
If one does not enforce this asymmetry as an initial condition and consider the Universe to start in a matter antimatter symmetric manner, then one has to satisfy the Sakharov's conditions \cite{sakharov} in order to produce a net baryon asymmetry. These conditions namely, (i) baryon number (B) violation, (ii) C and CP violation, (iii) departure from thermal equilibrium can in principle, be satisfied within a particle physics framework. However, the third condition can not be satisfied within the framework of SM of particle physics for the observed value of the Higgs boson mass and the observed CP violation in the quark sector is way too small to generate the observed baryon asymmetry. Thus, one has to go beyond the SM in order to generate the observed baryon asymmetry in a symmetric Universe to begin with.

Several BSM proposals have been proposed and well studied in the literature to solve either one or multiple of the three problems mentioned above. For example, the seesaw mechanism \cite{ti} in order to generate tiny Majorana masses of neutrinos can also address the baryon asymmetry problem within the leptogenesis framework (For a review of leptogenesis, one can refer to the review article \cite{davidsonPR}). Within this framework, the observed baryon asymmetry is generated by creating a leptonic asymmetry first and then converting it into baryon asymmetry through $B+L$ violating electroweak sphaleron transitions \cite{sphaleron}. According to the original proposal of Fukugita and Yanagida \cite{fukuyana}, the out of equilibrium CP violating decay of heavy right handed neutrinos present in the type I seesaw mechanism can naturally produce the required lepton asymmetry. Although the issue of dark matter remain disconnected from this minimal setup, several interesting proposals have appeared which relate DM with the mechanism of baryon asymmetry and (or) neutrino mass. For example, the scotogenic models \cite{ma06} related DM and neutrino mass, WIMPy baryogenesis (leptogenesis) models \cite{wimpy} and asymmetric dark matter models \cite{ADM} relate DM with the origin of baryon asymmetry etc.

Another equally interesting but much less explored scenario for neutrino mass and baryon asymmetry is the so called Dirac leptogenesis, first proposed by \cite{diraclepto0} and later extended to realistic phenomenological models by \cite{diraclepto1, diraclepto10, diraclepto2, diraclepto20, diraclepto3}. These scenarios assume the light neutrinos to be of Dirac nature contrary to the usual Majorana light neutrinos in conventional seesaw models. In these scenarios, the total lepton number or rather $B-L$ is conserved just like in the SM and hence there exists no net lepton asymmetry created. However, if one can create an equal and opposite amount of lepton asymmetry in the left and right handed sectors, the electroweak sphalerons can later convert the lepton asymmetry in the left handed sector into a net baryon asymmetry. The lepton asymmetries left and right handed sectors are prevented from equilibration due to the tiny effective Dirac Yukawa couplings. Although several different models are there in the literature that can generate tiny Dirac neutrino masses \cite{diracmass, diracmass1}, we focus on scotogenic type Dirac neutrino mass models \cite{ma1} in order to accommodate dark matter naturally into the model. Like most particle physics models of Dirac neutrinos, we also consider additional symmetries to forbid the Majorana masses of gauge singlet fermions and stabilise the dark matter candidate simultaneously. For the purpose of minimality, we consider a discrete symmetry $Z_4 \times Z_3$ whereas the new physics sector consists of two different types of gauge singlet fermions $\psi, \nu_R$, three additional scalar fields $\phi_2, \eta, \chi$ apart from the SM particle content. The neutrino mass arise at one loop level and the decay of heavy fermion $\psi$ creates the required asymmetries in the left as well as right handed sectors. The lighter of the scalar fields $\phi_2, \chi$ can be a DM candidate in the model, after considering the relevant bounds on dark matter relic density, direct detection cross section as well as invisible decay width of the SM Higgs boson. We also calculate the new physics contribution to the charged lepton flavour violating (LFV) decay process $(\mu \rightarrow e \gamma)$ and show it to remain within reach of current experimental sensitivity \cite{MEG16} for the region of parameter space allowed from the requirement of successful Dirac leptogenesis. We also take into account the new physics contribution to the electric dipole moment (EDM) of charged leptons. Since the model predicts zero lepton number violation (LNV) and hence no $0\nu \beta \beta $, any future observation of this LNV process will also partially rule out Dirac leptogenesis as the only source of baryon asymmetry due to the presence of additional source of creating lepton asymmetry.

This paper is organized as follows. In section \ref{model}, we discuss our model of scotogenic Dirac neutrino mass and then discuss the generation of tiny neutrino mass at one-loop level in section \ref{numass}. In section \ref{dm}, we briefly discuss calculation of dark matter relic abundance as well as direct detection cross section. In section \ref{lepto}, we outline the calculation of baryon asymmetry through Dirac leptogenesis and then discuss charged lepton flavour violation in section \ref{lfv}. We finally discuss our results and conclude in section \ref{conclude}.
\begin{center}
\begin{table}
\caption{Particle Content of the Model}
\begin{tabular}{|c|c||c|}
\hline
Particle & $SU(3)_c \times SU(2)_L \times U(1)_Y$ & $Z_4 \times Z_3$ \\
\hline
$ Q=(u,d)_L $ & $(3,2,\frac{1}{6})$  & $(1,1)$ \\
$ u_R $ & $(\bar{3},1,\frac{2}{3})$  & $(1,1)$ \\
$ d_R $ & $(\bar{3},1,-\frac{1}{3})$  & $(1,1)$\\
$L= (\nu, e)_L $ & $(1,2,-\frac{1}{2})$ & $(1,1)$ \\
$e_R$ & $(1,1,-1)$ & $(1,1)$ \\
\hline
$\nu_R$ & $(1,1,0)$ & $(1,\omega)$ \\
$\psi_L$ & $(1,1,0)$ & $(i, 1)$ \\
$\psi_R$ & $(1,1,0)$ & $(i, 1)$ \\
\hline
$ \phi_1=(\phi^+,\phi^0)_1 $ & $(1,2,-\frac{1}{2})$ & $(1, 1)$ \\
$\phi_2= (\phi^+,\phi^0)_2 $ & $(1,2,-\frac{1}{2})$ & $(-i, 1)$ \\
$\chi$ & $(1,1,0)$ & $(-i, \omega)$ \\
$\eta$ & $(1,1,0)$ & $(1, \omega)$ \\

\hline
\end{tabular}
\label{table1}
\end{table}
\end{center}
\section{The Model}
\label{model}
If we extend the SM just by three copies of right handed neutrinos $\nu_R$, there arises two difficulties in generating tiny Dirac neutrino masses: (i) the relevant Dirac Yukawa couplings have to be unnaturally fine tuned to $Y_{\nu} \leq 10^{-12}$ and (ii) the symmetry of the SM does not prevent Majorana mass term of singlet neutrinos $\nu_R$. Even if we forbid the Majorana mass term of $\nu_R$ by invoking the presence of additional symmetries like it was done in several earlier works \cite{diracmass}, we still have to fine tune the Dirac Yukawa couplings. One can also consider a second Higgs doublet, which acquires a tiny vacuum expectation value (vev) so that even order one Yukawa couplings can generate tiny Dirac neutrino masses \cite{diracmass1}. Instead of choosing tiny vev or tiny Yukawa couplings, here we try to generate such a small coupling at one loop level by introducing additional particles which take part in the loop. The model which we consider has the particle content as shown in table \ref{table1}. The symmetry of the model is an extension of the standard model gauge symmetry by a discrete symmetry $Z_4 \times Z_3$. The additional discrete symmetry $Z_4$ is chosen in order to prevent Majorana masses of $\psi$ as well as to stabilise the DM candidate going inside the loop. The $Z_3$ symmetry is chosen to prevent the Majorana mass term of the right handed neutrino $\nu_R$.  The earlier work on one loop Dirac neutrino masses \cite{ma1} considered $U(1)_{B-L} \times Z_2 \times Z_2$ as the additional symmetry. For the sake of minimality, here we stick to a setup with a $Z_4 \times Z_3$ extension of the SM symmetries. The Yukawa Lagrangian of the model can be written as 
\begin{align}
\mathcal{L} & = Y_u \overline{Q} \phi_1 u_R + Y_d \overline{Q} \phi^{\dagger}_1 d_R + Y_e \overline{L} \phi^{\dagger}_1 e_R+ h \overline{L} \phi_2 \psi_R+ M_\psi \overline{\psi_L} \psi_R + h^{'} \overline{\nu_R} \chi \psi_L +\text{h.c.}
\end{align}
The neutral component of the doublet scalar $\phi_1$ and the singlet $\eta$ acquire a non-zero vacuum expectation value (vev) to break the symmetry of the model as
$$ SU(3)_c \times SU(2)_L  \times U(1)_{Y} \times Z_4 \times Z_3  \quad \underrightarrow{\langle
\phi^0_1, \eta \rangle} \quad SU(3)_c\times U(1)_{\text{em}} \times Z_4 $$
The remnant $Z_4$ symmetry stabilises the dark matter particle whereas $Z_3$ gets spontaneously broken in order to allow the one loop mixing of $\nu_L$ and $\nu_R$. The quarks and charged leptons acquire their masses from the vev of $\phi^0_1$ whereas the neutrinos remain massless at tree level. The dark matter in this model is the lighter scalar among $\phi^0_2$ and $\chi$, as we are considering the Dirac fermion $\psi$ to be much heavier from leptogenesis point of view, as we discuss in the remaining sections in further details.

\begin{figure}[htb]
\centering
\includegraphics[scale=0.75]{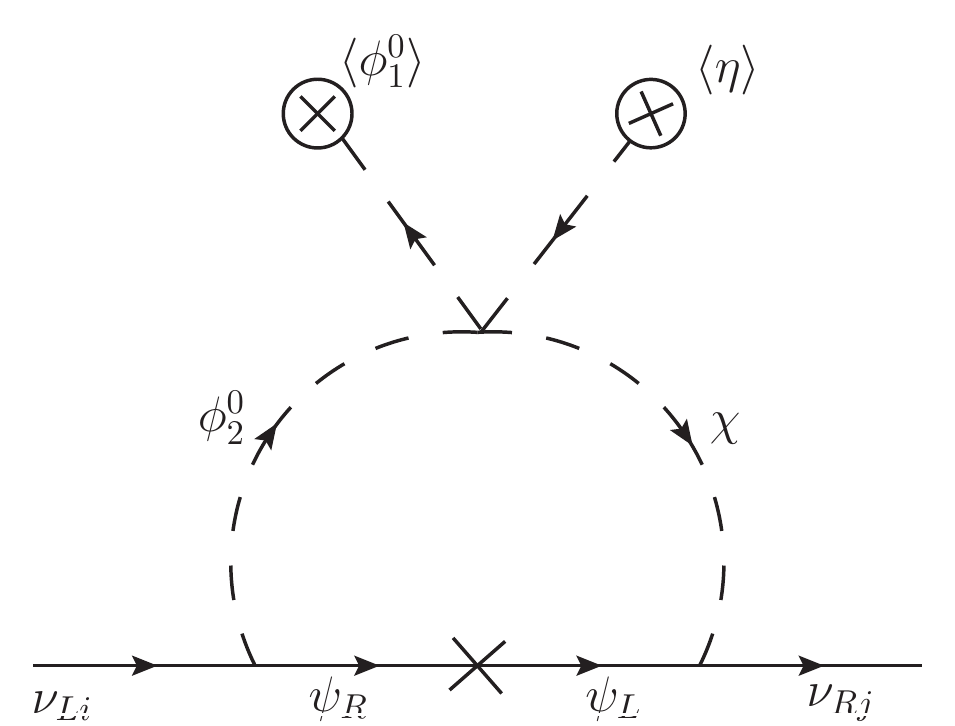}
\caption{One-loop contribution to Dirac neutrino mass}
\label{numass}
\end{figure}
\section{Dirac Neutrino Mass}
\label{numass}
The neutrinos, which remain massless at tree level, acquire a Dirac mass at one loop level as shown by the Feynman diagram in figure \ref{numass}. The calculation of scotogenic Dirac mass at one loop, first shown in \cite{ma1} follows a similar procedure as in the one loop scotogenic Majorana masses \cite{ma06}. Let us consider the quartic term $\lambda_4 \phi^{\dagger}_1 \chi^{\dagger} \phi_2 \eta$ with real $\lambda_4$. Without this term, there is no mixing between $\phi^0_2, \chi$ and hence the diagram shown in figure \ref{numass} gives vanishing contribution to Dirac neutrino mass. Let us denote the vev's as $\langle \phi^0_1 \rangle = v/\sqrt{2}, \langle \eta \rangle = u/\sqrt{2}$ and the fields as $\phi^0_2 = (\phi^0_{2R} +i \phi^0_{2I})/\sqrt{2}, \chi = (\chi_R + i \chi_I)/\sqrt{2}$. Assuming no mixing between real and imaginary components of individual fields, the above quartic term will introduce a mixing between $\phi^0_{2R}, \chi_R$ as well as between $\phi^0_{2I}, \chi_I$. Let, $\xi_{1,2}$ be the mass eigenstates of $\phi^0_{2R}, \chi_R$ sector with a mixing angle $\theta_1$. Similarly,  $\zeta_{1,2}$ be the mass eigenstates of $\phi^0_{2I}, \chi_I$ sector with a mixing angle $\theta_2$. The contribution of the real sector $\phi^0_{2R}, \chi_R$ to one loop Dirac neutrino mass \cite{ma1} can then be written as 
\begin{equation}
(m_{\nu})_{Rij} =\frac{\sin{\theta_1} \cos{\theta_1}}{32 \pi^2} \sum_k h_{ik}h^{'}_{kj} M_{\psi k} \left ( \frac{m^2_{\xi_1}}{m^2_{\xi_1}-M^2_{\psi k}} \text{ln} \frac{m^2_{\xi_1}}{M^2_{\psi k}}-\frac{m^2_{\xi_2}}{m^2_{\xi_2}-M^2_{\psi k}} \text{ln} \frac{m^2_{\xi_2}}{M^2_{\psi k}} \right)
\label{numassR}
\end{equation}
Similarly one can write down the contribution $(m_{\nu})_{Iij}$ from the imaginary sector $\phi^0_{2I}, \chi_I$. The total neutrino mass is 
\begin{equation}
(m_{\nu})_{ij} = (m_{\nu})_{Rij}+(m_{\nu})_{Iij}
\end{equation}
In the absence of any fine-tuned cancellations between different terms in the neutrino mass formula above, we expect each of the terms to be of sub-eV scale. Assuming $m_{\zeta_1} = 100$ GeV and $M_{\psi} = 10$ TeV, the first term in the expression becomes 
$$ (m_{\nu})^{1}_{Rij} = 1.46 \times 10^{-2} \sin{2 \theta_1} \sum_k h_{ik}h^{'}_{kj} $$
This can be at the sub-eV scale if 
\begin{equation}
\sin{2 \theta_1} h_{ik}h^{'}_{kj} < 10^{-8} 
\label{cond1}
\end{equation}
This condition can be easily satisfied by suitable tuning of the Yukawa couplings $h. h^{'}$. The mixing angles $\theta_1, \theta_2$ can be fixed by choosing the mass terms, vev's and relevant couplings as
$$\tan{2 \theta_1} = \frac{\lambda_4 v u}{m^2_{\chi_R}-m^2_{\phi^0_{2R}}}, \;\;\;\; \tan{2 \theta_2} = \frac{\lambda_4 v u}{m^2_{\chi_I}-m^2_{\phi^0_{2I}}}$$
where $m^2_{\chi_{R,I}}, m^2_{\phi^0_{2R, 2I}}$ are related to the mass eigenvalues as 
$$ m^2_{\xi_1} = m^2_{\phi^0_{2R}} +\frac{\lambda_4 u v}{4(m^2_{\phi^0_{2R}}-m^2_{\chi_{R}})}, \;\; m^2_{\xi_2} = m^2_{\chi_{R}} -\frac{\lambda_4 u v}{4(m^2_{\phi^0_{2R}}-m^2_{\chi_{R}})} $$
$$ m^2_{\zeta_1} = m^2_{\phi^0_{2I}} +\frac{\lambda_4 u v}{4(m^2_{\phi^0_{2I}}-m^2_{\chi_{I}})}, \;\; m^2_{\zeta_2} = m^2_{\chi_{I}} -\frac{\lambda_4 u v}{4(m^2_{\phi^0_{2I}}-m^2_{\chi_{I}})} $$
Here, we have assumed $\lambda_4 u v \ll m^2_{\phi^0_{2R}}-m^2_{\chi_{R}}, m^2_{\phi^0_{2I}}-m^2_{\chi_{I}}$.
Thus, choosing different mass terms and dimensionless couplings of the scalar Lagrangian appropriately leads to the required mixing angles $\theta_{1,2}$ in order to be in agreement with neutrino mass scale \eqref{cond1}.

\section{Dark Matter}
\label{dm}
The relic abundance of a dark matter particle $\chi$ which was in thermal equilibrium at some earlier epoch can be calculated by solving the Boltzmann equation
\begin{equation}
\frac{dn_{\chi}}{dt}+3Hn_{\chi} = -\langle \sigma v \rangle (n^2_{\chi} -(n^{\text{eqb}}_{\chi})^2)
\end{equation}
where $n_{\chi}$ is the number density of the dark matter particle $\chi$ and $n^{eqb}_{\chi}$ is the number density when $\chi$ was in thermal equilibrium. $H$ is the Hubble expansion rate of the Universe and $ \langle \sigma v \rangle $ is the thermally averaged annihilation cross section of the dark matter particle $\chi$. In terms of partial wave expansion $ \langle \sigma v \rangle = a +b v^2$. Clearly, in the case of thermal equilibrium $n_{\chi}=n^{\text{eqb}}_{\chi}$, the number density is decreasing only by the expansion rate $H$ of the Universe. The approximate analytical solution of the above Boltzmann equation gives \cite{Kolb:1990vq, kolbnturner}
\begin{equation}
\Omega_{\chi} h^2 \approx \frac{1.04 \times 10^9 x_F}{M_{Pl} \sqrt{g_*} (a+3b/x_F)}
\end{equation}
where $x_F = m_{\chi}/T_F$, $T_F$ is the freeze-out temperature, $g_*$ is the number of relativistic degrees of freedom at the time of freeze-out and $M_{Pl} \approx 10^{19}$ GeV is the Planck mass. Here, $x_F$ can be calculated from the iterative relation 
\begin{equation}
x_F = \ln \frac{0.038gM_{\text{Pl}}m_{\chi}<\sigma v>}{g_*^{1/2}x_F^{1/2}}
\label{xf}
\end{equation}
The expression for relic density also has a more simplified form given as \cite{Jungman:1995df}
\begin{equation}
\Omega_{\chi} h^2 \approx \frac{3 \times 10^{-27} \text{cm}^3 \text{s}^{-1}}{\langle \sigma v \rangle}
\label{eq:relic}
\end{equation}
The thermal averaged annihilation cross section $\langle \sigma v \rangle$ is given by \cite{Gondolo:1990dk}
\begin{equation}
\langle \sigma v \rangle = \frac{1}{8m^4_{\chi}T K^2_2(m_{\chi}/T)} \int^{\infty}_{4m^2_{\chi}}\sigma (s-4m^2_{\chi})\surd{s}K_1(\surd{s}/T) ds
\label{eq:sigmav}
\end{equation}
where $K_i$'s are modified Bessel functions of order $i$, $m_{\chi}$ is the mass of Dark Matter particle and $T$ is the temperature.

If we consider the neutral component of the scalar doublet $\phi_2$ to be the dark matter candidate, the details of relic abundance calculation is similar to the inert doublet model studied extensively in the literature \cite{ma06,Barbieri:2006dq,Majumdar:2006nt,LopezHonorez:2006gr,ictp,borahcline, honorez1,DBAD14}. In the low mass regime $m_{DM} \leq M_W$, dark matter annihilation into the SM fermions through s-channel Higgs mediation dominates over other channels. Beyond the W boson mass threshold, the annihilation channel of scalar doublet dark matter into $W^+W^-$ pairs opens up suppressing the relic abundance below what is observed by Planck experiment, unless the dark matter mass is heavier than around 500 GeV, depending on the DM-Higgs coupling. Apart from the usual annihilation channels of inert doublet dark matter, in this model there is another interesting annihilation channel where dark matter annihilates into $\nu \bar{\nu}$ through the heavy fermion $\psi$ in the t-channel. As we discuss later, this annihilation channel however remains suppressed compared to others for the choices of couplings and mass of $\psi$ in from successful leptogenesis criteria.

Apart from the relic abundance constraints from Planck experiment, there exists strict bounds on the dark matter nucleon cross section from direct detection experiments like Xenon100 \cite{Aprile:2013doa} and more recently LUX \cite{LUX, LUX16}. For scalar dark matter considered in this work, the relevant spin independent scattering cross section mediated by SM Higgs is given as \cite{Barbieri:2006dq}
\begin{equation}
 \sigma_{\text{SI}} = \frac{\lambda^2 f^2}{4\pi}\frac{\mu^2 m^2_n}{m^4_h m^2_{DM}}
\label{sigma_dd}
\end{equation}
where $\mu = m_n m_{DM}/(m_n+m_{DM})$ is the DM-nucleon reduced mass and $\lambda$ is the quartic coupling involved in DM-Higgs interaction. A recent estimate of the Higgs-nucleon coupling $f$ gives $f = 0.32$ \cite{Giedt:2009mr} although the full range of allowed values is $f=0.26-0.63$ \cite{mambrini}. The latest LUX bound \cite{LUX16} on $\sigma_{\text{SI}}$ will constrain the DM-Higgs coupling $\lambda$ as we discuss in details in section \ref{conclude}. One can also constrain the DM-Higgs coupling $\lambda$ from the latest LHC constraint on the invisible decay width of the SM Higgs boson. This constraint is applicable only for dark matter mass $m_{DM} < m_h/2$. The invisible decay width is given by
\begin{equation}
\Gamma (h \rightarrow \text{Invisible})= {\lambda^2 v^2\over 64 \pi m_h} 
\sqrt{1-4\,m^2_{DM}/m^2_h}
\end{equation}
The latest ATLAS constraint on invisible Higgs decay is \cite{ATLASinv}
$$\text{BR} (h \rightarrow \text{Invisible}) = \frac{\Gamma (h \rightarrow \text{Invisible})}{\Gamma (h \rightarrow \text{Invisible}) + \Gamma (h \rightarrow \text{SM})} < 22 \%$$
As we will discuss in the section \ref{conclude}, this bound is weaker than the conservative LUX 2016 bound.

\begin{figure}[htb]
\centering
\includegraphics[scale=0.75]{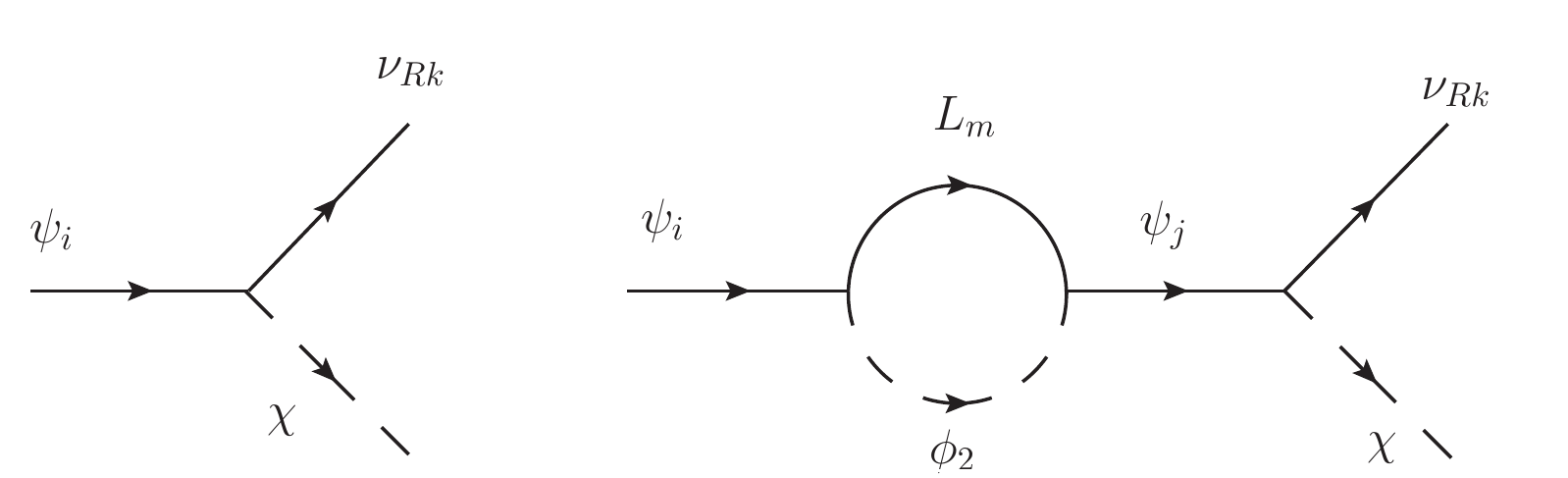}
\caption{Decay of heavy singlet fermion $\psi$ into $\nu_R$ and $\chi$}
\label{leptofig}
\end{figure}

\section{Dirac Leptogenesis}
\label{lepto}
Here we revisit the idea of Dirac leptogenesis studied in the works \cite{diraclepto0, diraclepto1, diraclepto10, diraclepto2, diraclepto20, diraclepto3} and particularly adopt the setup outlined in \cite{diraclepto2} for the calculation of baryon asymmetry. In our model, the heavy singlet Dirac fermion $\psi = \psi_L + \psi_R$ can decay into $L \phi_2$ as well as $\nu_R \chi$ to create lepton asymmetry in the left and right handed lepton sectors. Since there is no total lepton number violation, hence the asymmetry created in the left and right handed sectors are equal and opposite to each other $\epsilon_{\text{Total}} = \epsilon_L + \epsilon_R = 0$. Due to tiny one loop suppressed effective couplings between the left and right handed neutrino sectors, these corresponding asymmetries in the two sectors do not get equilibrated upto a very low temperature, below the electroweak transition. The sphalerons during the electroweak phase transition therefore, can convert the non-zero asymmetry in the left handed sector into a net baryon asymmetry, which survive afterwards to give rise to the observed baryon asymmetry of the Universe. Due to the singlet nature of right handed neutrinos under the electroweak gauge group, the electroweak sphalerons can not convert the asymmetry stored in the right handed sector into a baron asymmetry, as long as tiny effective coupling between left and right handed neutrinos prevent them from equilibrating. The conversion between $B-L$ asymmetry into a net baryon (B) asymmetry by electroweak sphalerons is given by the relation \cite{BLsphaleronB}
\begin{equation}
Y_B = \frac{28}{29} Y_{B-L}
\end{equation} 
If leptogenesis ends before the sphaleron processes become active $(T \geq 10^{12} \; \text{GeV})$, then $Y_{B-L} = -Y_L$. Since the asymmetry in left and right handed sectors have equal magnitudes, it is sufficient to calculate the asymmetry in one sector. Thus, we can write 
$$ Y_B = -\frac{28}{29} Y_{L_{\nu_R}} $$
similar to the way it was done by \cite{diraclepto2}. The asymmetry in the right handed sector can be generated by the decay of $\psi$ through the Feynman diagrams shown in figure \ref{leptofig}. The Boltzmann equation for Dirac leptogenesis for the decay of $\psi_i$ can be written similar to the ones given in \cite{diraclepto2} as
\begin{align}
\frac{d \psi_{\Sigma \psi_1}}{dz} &= \frac{z}{H(z=1)}\left[2 - 
\frac{\eta_{\Sigma \psi_1} }{\eta^{eq}_{\psi_1}} + \delta_R 
\left(3\frac{\eta_{\Delta L}}{2} + \eta_{\Delta \psi_1}\right)\right]\Gamma^D \nonumber \\
 \frac{d \psi_{\Delta \psi_1}}{dz} &= \frac{z}{H(z=1)} \left[\eta_{\Delta L} - \frac{\eta_{\Delta \psi_1} }{\eta^{eq}_{\psi_1}} + B_R \left(3 \frac{\eta_{\Delta L}}{2} + \eta_{\Delta \psi_1}\right)\right]\Gamma^D \nonumber \\
 \frac{d \psi_{\Delta L}}{dz} &= \frac{z}{H(z=1)} \left\{\left[\delta_R \left(1 - \frac{\eta_{\Sigma \psi_1}}{2\eta^{eq}_{\psi_1} }\right) - \left(1 - \frac{B_R}{2}\right)\left(\eta_{\Delta L} - \frac{\eta_{\Delta \psi_1}}{\eta^{eq}_{\psi_1} }\right) \right]\Gamma^D  \right. \nonumber \\ 
 &- \left. \left(3\frac{\eta_{\Delta L}}{2} 
 + \eta_{\Delta \psi_1}\right)\Gamma^W \right\}
\end{align}
where $\eta_{\Sigma \psi_1} = \frac{\eta_{\psi_1} + \overline{\eta}_{\psi_1}  }{\eta_\gamma}$, $\eta_{\Delta \psi} = \frac{\eta_{\psi_1} - \overline{\eta}_{\psi_1}  }{\eta_\gamma}$, $z=M_{\psi_1}/T$ and 
\begin{align*}
\Gamma^D &= \left( \Gamma(\psi_1 \rightarrow \nu_R \chi) + \Gamma(\psi_1 \rightarrow L \phi_2) \right)\frac{z^2}{2}K_1(z) = \Gamma^W
\end{align*}
with $K_1 (z)$ being the first order modified Bessel function. These equations include the decay and inverse decays of $\psi_1$ as well as the CP violating $2 \leftrightarrow 2$ scattering contributions. In the above equations, $\Gamma^W$ includes the wash-out processes involving higher powers of Yukawa couplings, which tend to equilibrate the asymmetries created in the left and right handed neutrino sectors. These are $L\phi_2 \leftrightarrow \chi \nu_R$ mediated by the s-channel exchange of $\psi_i$, $L\bar{\nu}_R \leftrightarrow \phi_2 \chi$ mediated by the t-channel exchange of $\psi_i$. These processes at high temperature can be approximated to be
\begin{align}
\Gamma_{L-R} &\sim \frac{|h|^2|h'|^2}{M^4_1}T^5
\end{align}
which in turn should be less than the Hubble expansion rate at the radiation dominated era given as
\begin{align}
H(T) &= \sqrt[]{\frac{8\pi^2g_*}{90}}\frac{T^2}{M_{Pl}}
\end{align}
The strongest bounds comes from the high temperature when the asymmetry is produced i.e $z \simeq 1$
\begin{align}
\frac{|h|^2|h'|^2}{M_1} &\leq \frac{1}{M_p}\sqrt[]{\frac{8\pi^2 g_*}{90}}
\label{outofeqb}
\end{align}
However, the dominant contribution would come from decay and inverse decay for the left-right equilibrium and also from the subsequent decay of a real $\psi_i$ or $\bar{\psi}_i$. Since in our case we have reasonably large hierarchy in mass of $\psi_i$'s, the contribution from the off shell processes $L \phi_2 \leftrightarrow \nu_R \chi$ is bounded from above by $\Gamma^D$ for the region around $z\sim 1$. Therefore, we have considered a conservative approximation that 
$\Gamma^W = \Gamma^D$ similar to \cite{diraclepto2}. We solve these Boltzmann equations to determine the baryon asymmetry as discussed below.
\begin{figure}[htb]
\centering
\includegraphics[scale=0.75]{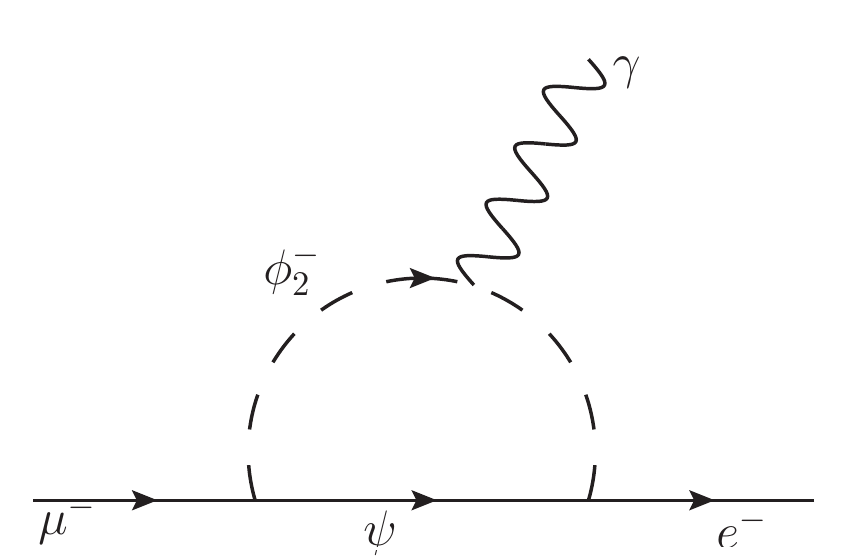}
\caption{One loop contribution to $\mu \rightarrow e \gamma$}
\label{lfv}
\end{figure}


\begin{figure}[htb]
\centering
\includegraphics[scale=0.5]{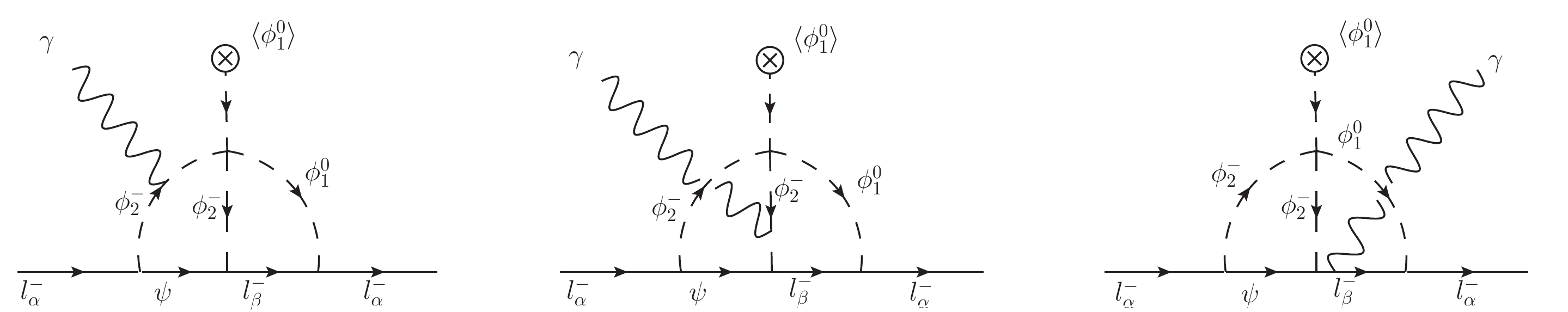}
\caption{Two loop contribution to electric dipole moment of charged leptons}
\label{lfv3}
\end{figure}

\begin{figure}
\centering
\begin{tabular}{c}
\epsfig{file=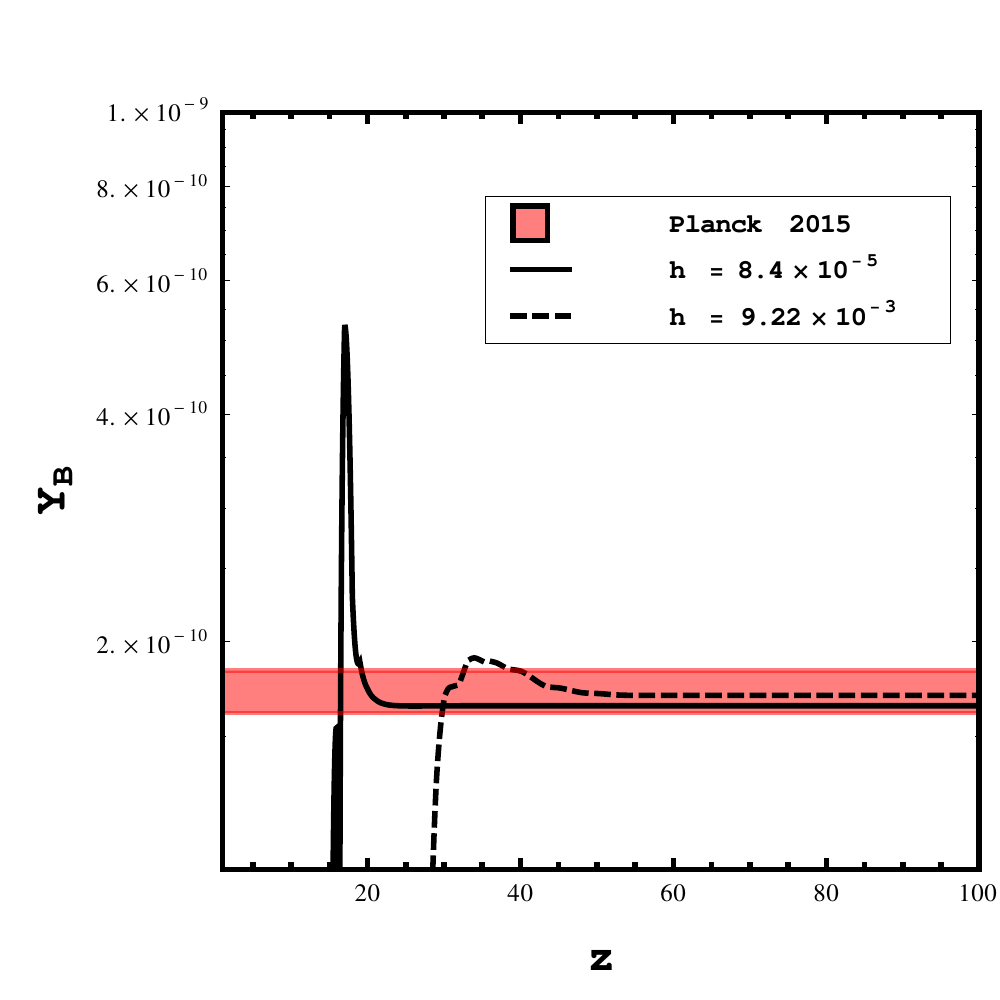,scale=1.0,clip=}
\end{tabular}
\caption{Baryon asymmetry as a function of $z=M_{\psi_1}/T$ for two different values of Yukawa couplings $h$. The horizontal band corresponds to the Planck 2015 constraint on baryon asymmetry \eqref{barasym}.}
\label{figlepto}
\end{figure}

\begin{figure}
\centering
\begin{tabular}{cc}
\epsfig{file=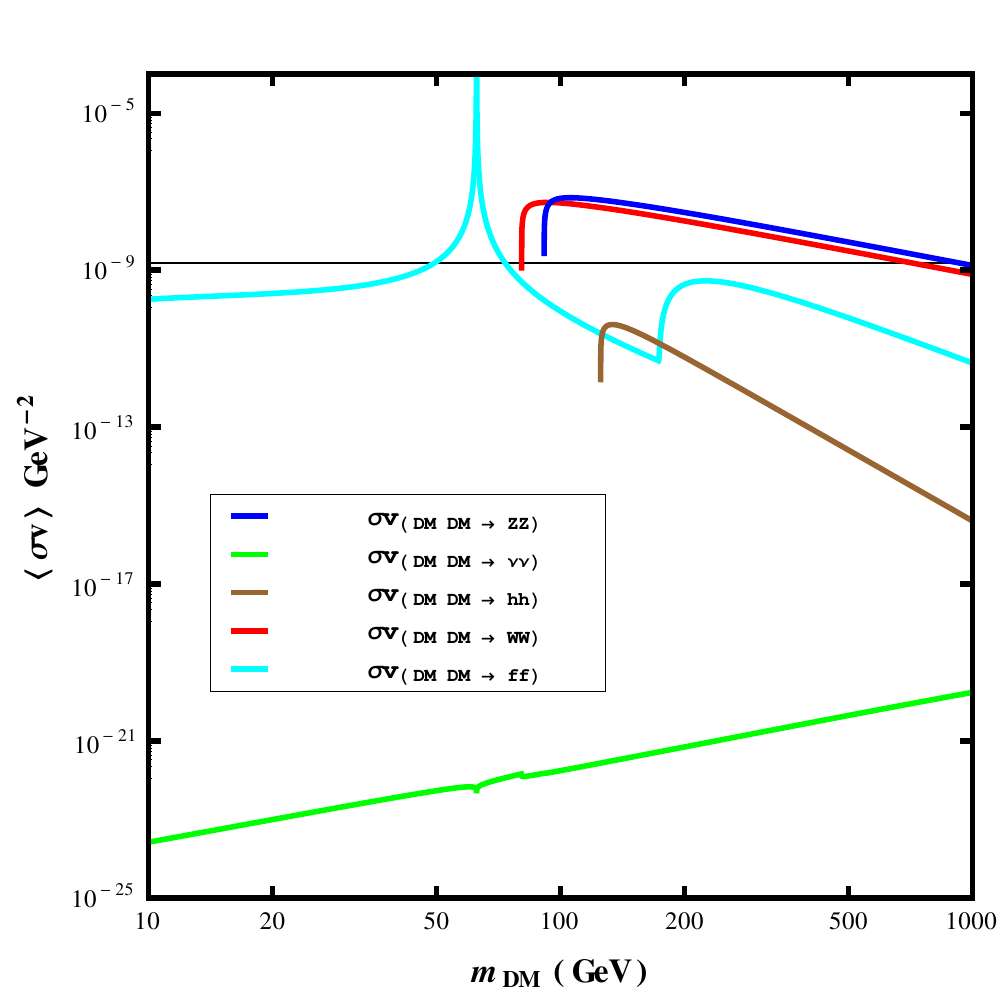,scale=0.7,clip=} &
\epsfig{file=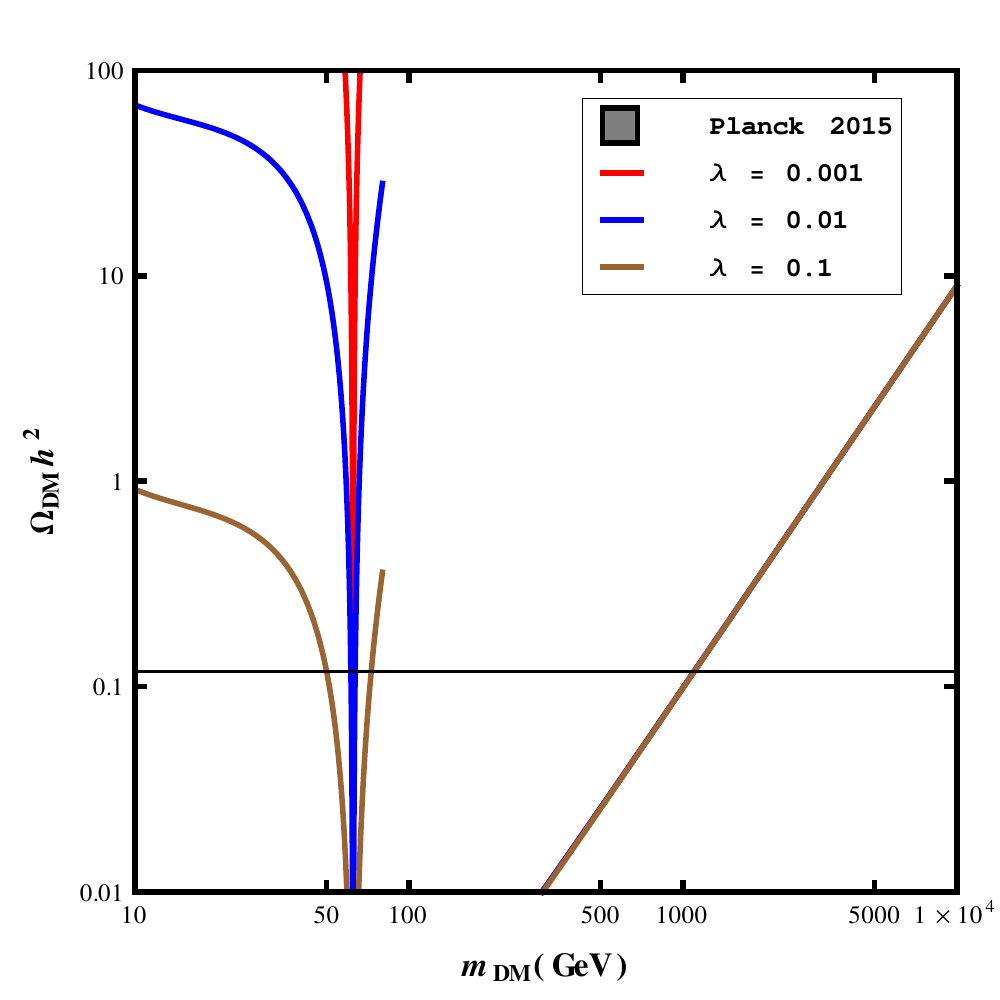,scale=0.7,clip=}
\end{tabular}
\caption{Left panel: Variation of different dark matter annihilation channels as a function of dark matter mass. The horizontal line corresponds to the typical annihilation cross section (from \eqref{eq:relic}) required to generate the correct relic abundance \eqref{dm_relic}. Right panel: The relic abundance of dark matter as a function of dark matter mass for three different values of DM-Higgs coupling $\lambda$ showing two different mass range satisfying relic density bound \eqref{dm_relic}.}
\label{figdm1}
\end{figure}

\begin{figure}
\centering
\begin{tabular}{c}
\epsfig{file=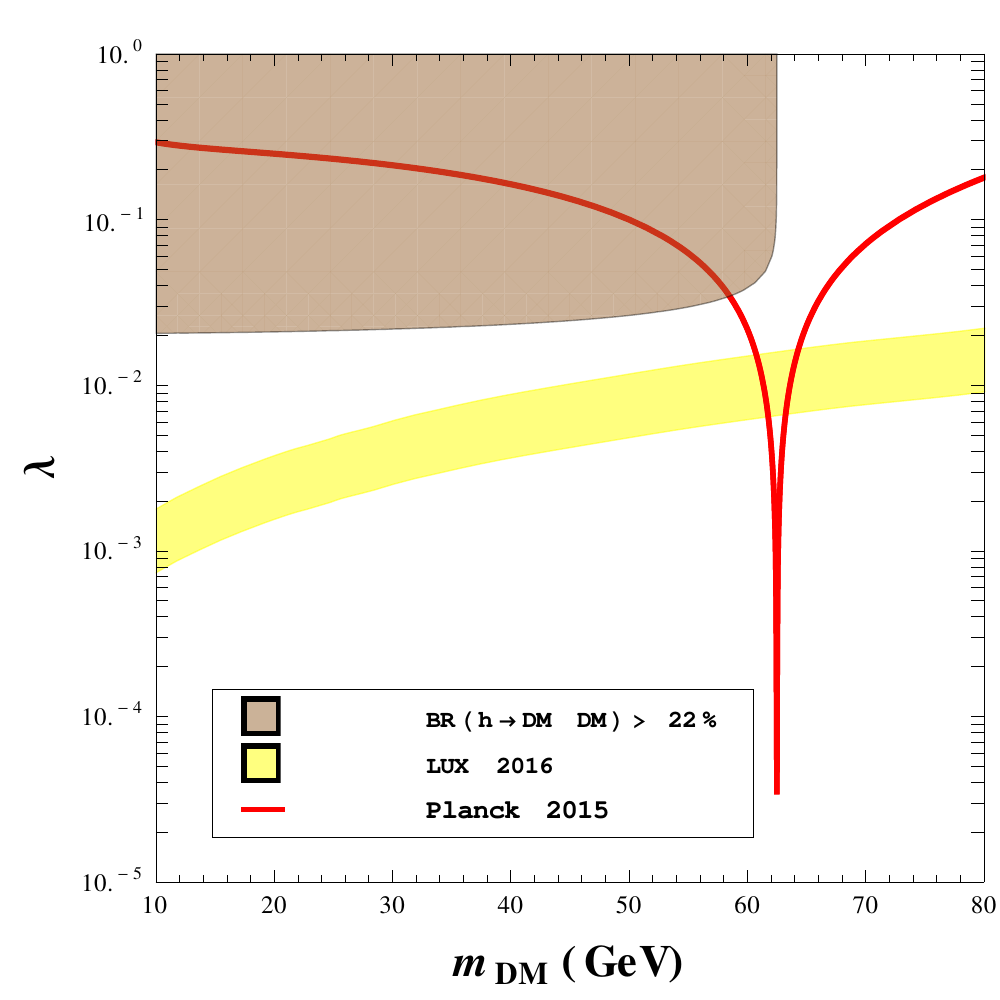,scale=1.0,clip=}
\end{tabular}
\caption{The parameter space in the $\lambda-m_{DM}$ plane for low mass regime $m_{DM} < M_W$ including constraints on relic abundance, direct detection as well as invisible Higgs decay. The final allowed region corresponds to the narrow region around resonance $m_{DM} = m_h/2$. }
\label{figdm2}
\end{figure}

\section{Flavor Violation and Electric Dipole Moment of Charged Leptons}
\label{lfv}
Charged lepton flavour violating decay is a promising process to study from BSM physics point of view. In the SM, such a process occurs at lone level and is suppressed by the smallness of neutrino masses, much beyond the current experimental sensitivity \cite{MEG16}. Therefore, any future observation of such LFV decays like $\mu \rightarrow e \gamma$ will definitely be a signature of new physics beyond the SM. In the present model, such new physics contribution can come from the charged component of the additional scalar doublet $\phi_2$. The relevant Feynman diagram is shown in figure \ref{lfv}. Adopting the general prescriptions given in \cite{LFV1}, the decay width of $\mu \rightarrow e \gamma$ can be calculated as
\begin{align}
\Gamma (\mu \rightarrow e \gamma) &= \frac{h^4 \left(m^2_\mu - m^2_e \right)^3(m^2_\mu + m^2_e)}{4096 \pi^5 m^3_\mu m^4_{\phi^-_2}} \left[\frac{ \left((t-1)(t(2t+5)-1) + 6t^2\ln t\right)^2}{144 (t-1)^8}\right]
\end{align}
where $t = m^2_{\psi_i}/m^2_{\phi^-_2}$. The corresponding branching ratio can be found by
$$ \text{BR}(\mu \rightarrow e \gamma) \approx \frac{\Gamma (\mu \rightarrow e \gamma)}{\Gamma_{\mu}} $$
where $\Gamma_{\mu} \approx 2.996 \times 10^{-19}$ GeV denotes the total decay width of muon. The latest bound from the MEG collaboration is $\text{BR}(\mu \rightarrow e \gamma) < 4.2 \times 10^{-13}$ at $90\%$ confidence level \cite{MEG16}.

On the other hand, electric dipole moment of charged leptons is a flavour conserving observable which is a measure of the particle's spin to an external electric field. In the SM, the EDM of the electron is generated only at four-loop level and hence is vanishingly small $\lvert d_e \rvert/e \sim 3 \times 10^{-38}$ cm \cite{EDM1}. This is way below the current experimental limits on the EDM of charged leptons:
\begin{equation}
\lvert d_e \rvert/e < 8.7 \times 10^{-29} \; \text{cm} \;\;\;\; (\text{ACME})
\end{equation}
\begin{equation}
\lvert d_{\mu} \rvert/e < 1.9 \times 10^{-19} \; \text{cm} \;\;\;\; (\text{Muon} \;g-2)
\end{equation}
\begin{equation}
\lvert \text{Re}(d_{\tau}) \rvert/e < 4.5 \times 10^{-17} \; \text{cm} \;\;\;\; (\text{Belle})
\end{equation}
\begin{equation}
\lvert \text{Re}(d_{\tau}) \rvert/e < 2.5 \times 10^{-17} \; \text{cm} \;\;\;\; (\text{Belle})
\end{equation}
which have been measured by the ACME collaboration \cite{EDM2}, the Muon $(g-2)$ collaboration \cite{EDM3} and the Belle collaboration \cite{EDM4} respectively. Since the bound on the electron EDM is very strong, new physics contribution can be expected to bring the electron EDM within the sensitivity of next generation ACME experiment. Since our model has additional scalars and fermions and hence new sources of CP violation, one can expect to have an enhanced EDM of electron. Models having two Higgs doublets with softly broken $Z_2$ discrete symmetry were studied from charged lepton EDM point of view in \cite{EDM5, EDM51}. For EDM calculations in other BSM scenarios, please refer to the review \cite{EDM1}. In the present model, we can have EDM at two-loop level shown in the diagram \ref{lfv3}. Since, the second Higgs doublet in our model is protected by an unbroken discrete symmetry, we do not have any one-loop EDM diagram as discussed in \cite{EDM5}. The expression for the EDM of electron from the new physics contribution through the Feynman diagrams shown in figure \ref{lfv3} can be given as
\begin{equation}
 d_l = 2\Im \left[\frac{Y_l \lambda_{\phi^-_2} v h^2 }{256 \pi^4} (\sum^2_{i=1}(\sigma^i_R 
	- \sigma^i_L)  )\right] 
\end{equation}
where the details of $\sigma_{L,R}$ are given in appendix \ref{appen1}.  Here $Y_l v = m_l$ is the mass of the charged lepton and $\lambda_{\phi^-_2}$ is the quartic coupling of Higgs to the charged component of the second Higgs doublet $\phi_2$. To evaluate the integrals involved in the above expression, we use the numerical techniques given by \cite{edmnum}.

Thus, lower the mass of $\phi^-_2$, more is the decay width and hence more probability of discovering it at ongoing and future experiments. The choice of $m_{\phi^-_2}$ must however, satisfy the existing experimental bounds. The LEP collider experiment data restrict the charged scalar mass to $m_{\phi^-_2} > 70-90$ GeV \cite{lep1}. Another important restriction on $m_{\phi^-_2}$ comes from the electroweak precision data (EWPD). Since the contribution of the additional doublet $\phi_2$ to electroweak S parameter is always small \cite{Barbieri:2006dq}, we only consider the contribution to the electroweak T parameter here. The relevant contribution is given by \cite{Barbieri:2006dq}
\begin{equation}
\Delta T = \frac{1}{16 \pi^2 \alpha v^2} [F(m_{\phi^-_2}, m_{A^0})+F(m_{\phi^-_2}, m_{H^0}) -F(m_{A^0}, m_{H^0})]
\end{equation}
where 
\begin{equation}
F(m_1, m_2) = \frac{m^2_1+m^2_2}{2}-\frac{m^2_1m^2_2}{m^2_1-m^2_2} \text{ln} \frac{m^2_1}{m^2_2}
\end{equation}
In this work, we take $m_{A^0} \approx m_{H^0} =m_{DM}$. The EWPD constraint on $\Delta T$ is given as \cite{honorez1}
\begin{equation}
-0.1 < \Delta T + T_h < 0.2
\end{equation}
where $T_h \approx -\frac{3}{8 \pi \cos^2{\theta_W}} \text{ln} \frac{m_h}{M_Z}$ is the SM Higgs contribution to the T parameter \cite{peskin}. Thus, choosing a particular value of DM mass will leave a particular window of $m_{\phi^-_2}$ from LEP and EWPD constraints. We calculate the $\text{BR}(\mu \rightarrow e \gamma)$ for these allowed values of $m_{\phi^-_2}$ as we discuss below.
\begin{figure}
\centering
\begin{tabular}{c}
\epsfig{file=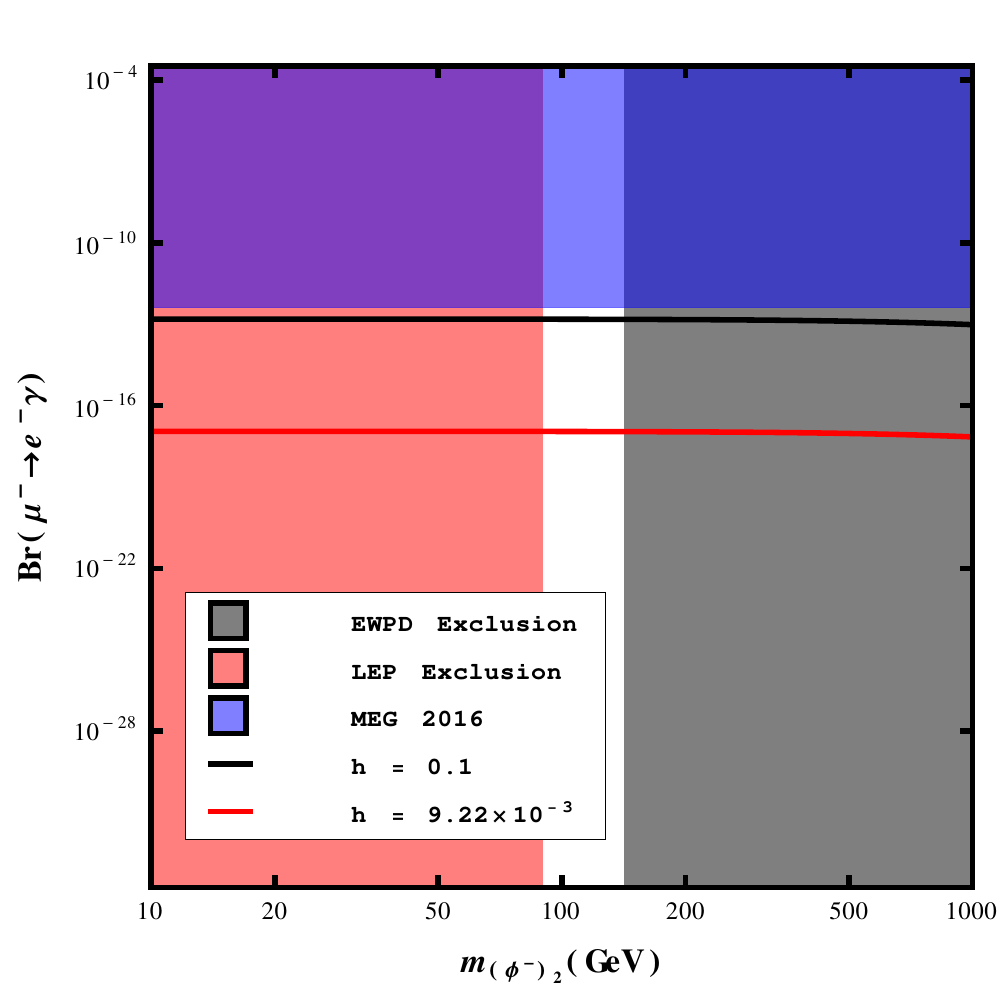,scale=1.0,clip=}
\end{tabular}
\caption{Contribution to $\mu \rightarrow e \gamma$ for two different values of Yukawa coupling $h$. The exclusion regions from MEG 2016 constraint, LEP constraint and EWPD constraints are also shown.}
\label{figlfv}
\end{figure}

\begin{figure}
\centering
\begin{tabular}{c}
\epsfig{file=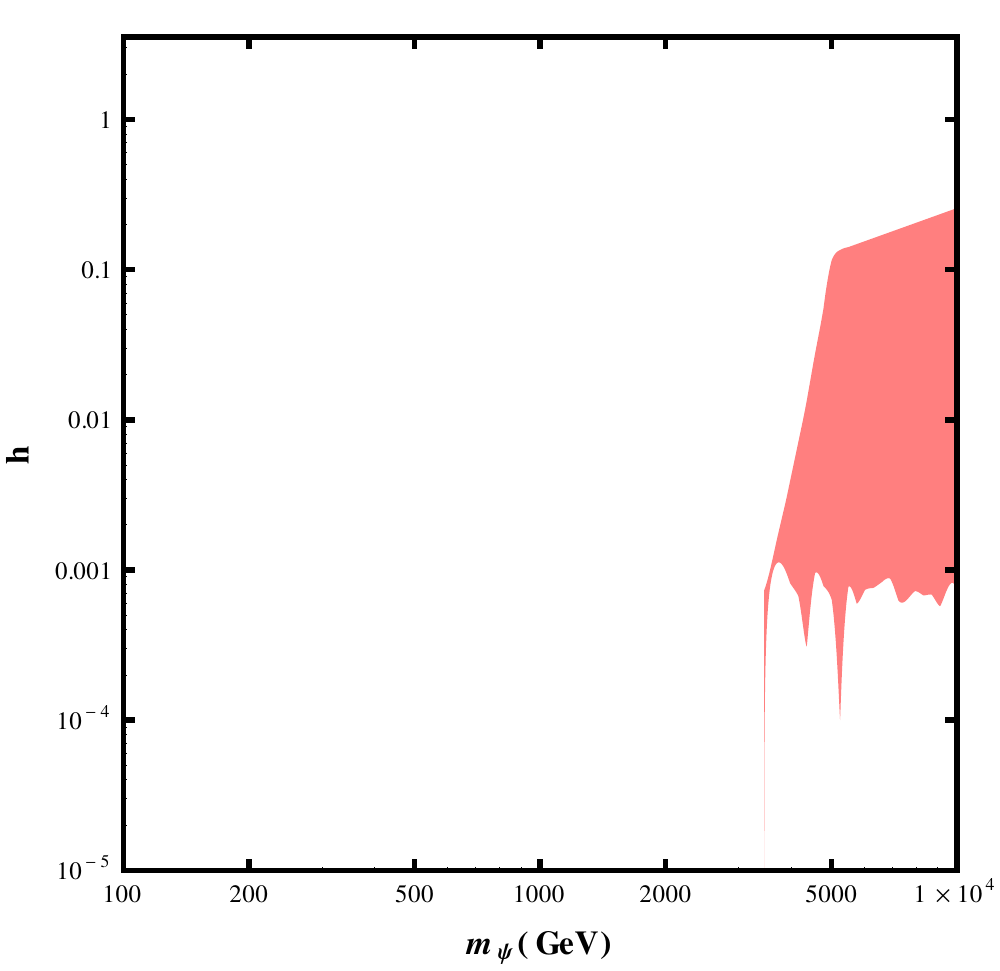,scale=1.0,clip=}
\end{tabular}
\caption{Allowed parameter space in $h-M_{\psi}$ plane from the requirement of successful leptogenesis, freeze-out of heavy fermion decay before sphaleron transitions and also incorporating the bounds on LFV decay $\mu \rightarrow e \gamma$. The masses charged and neutral component of $\phi_2$ are kept in the range that satisfy dark matter relic density as well as LEP bounds on electroweak parameters.}
\label{figlfv2}
\end{figure}

\begin{figure}
\centering
\begin{tabular}{c}
\epsfig{file=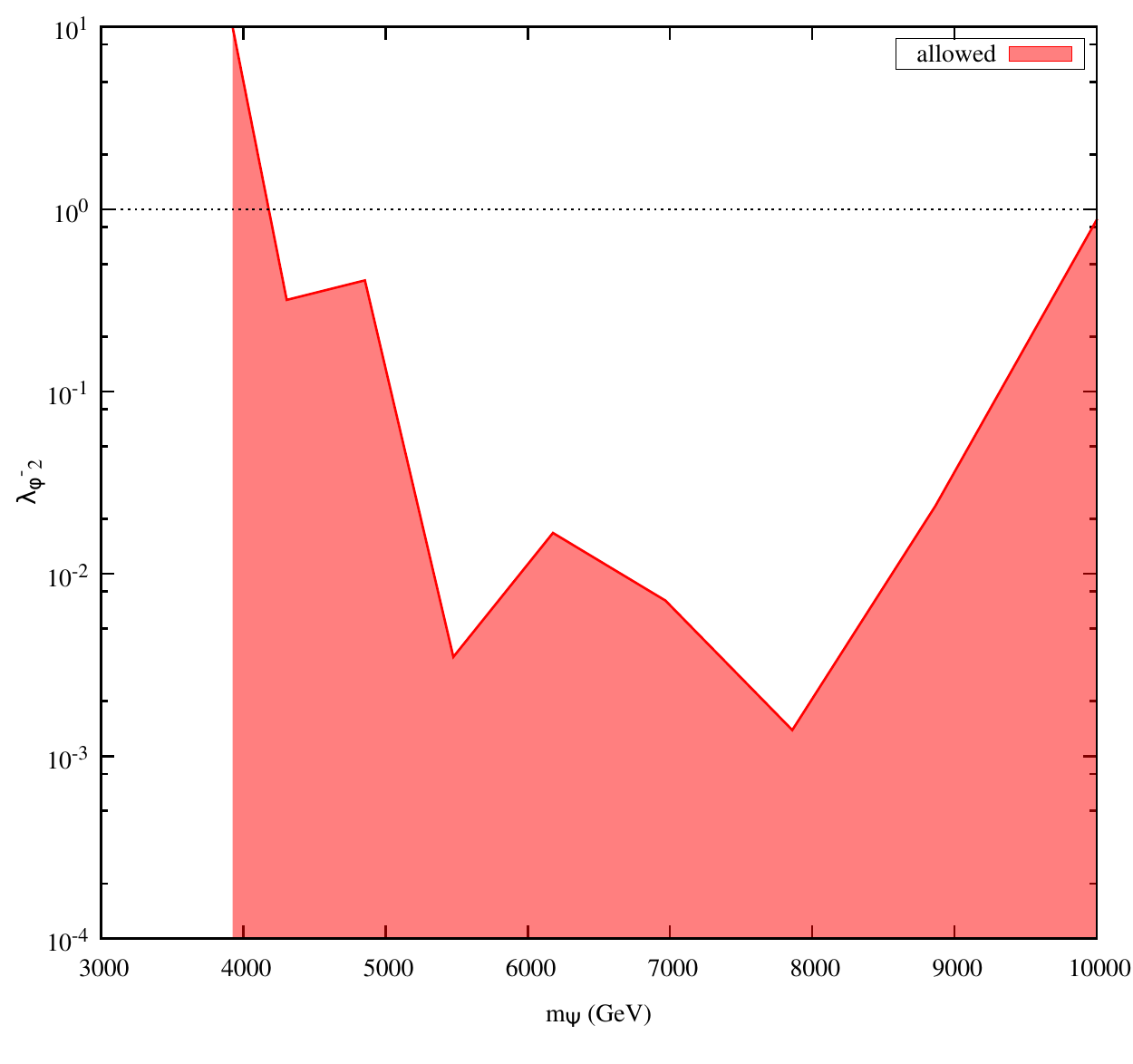,scale=1.0,clip=}
\end{tabular}
\caption{Allowed parameter space in $\lambda_{\phi^-_2}-M_{\psi}$ plane from the requirement of satisfying bounds on electron EDM, LFV decay $\mu \rightarrow e \gamma$ as well as the criteria of successful leptogenesis, freeze-out of heavy fermion decay before sphaleron transitions. The masses charged and neutral component of $\phi_2$ are kept in the range that satisfy dark matter relic density as well as LEP bounds on electroweak parameters.}
\label{figlfv3}
\end{figure}

\section{Results and Conclusion}
\label{conclude}
We have studied a minimal extension of the standard model in order to generate tiny Dirac neutrino masses at one loop level such that the particles going inside a loop can also generate the correct dark matter abundance and matter antimatter asymmetry in the Universe. Considering the heavy Dirac fermion decay as the source of lepton asymmetry, we write down the relevant Boltzmann equations and solve them numerically. Considering the lightest heavy fermion to have mass 5 TeV, the next to lightest one to be 12 TeV and equal Yukawa couplings $h=h^{'}$ , we calculate the final baryon asymmetry for different values of Yukawa couplings. The baryon asymmetry as a function of $z=M_{\psi_1}/T$ is shown in figure \ref{figlepto} for two different values of Yukawa couplings $h$. It can be seen from the figure that for both the values of Yukawa couplings, the final asymmetry $(z \rightarrow \infty)$ lies in the range observed by the Planck experiment. We also find the constraint on Yukawa coupling from the requirement that the heavy fermion decay process freezes out before the electroweak sphaleron temperature $T_{\text{sph}} \approx 140$ GeV and the lepton asymmetry gets converted into the correct baryon asymmetry observed. For the chosen masses of heavy fermions, this comes out to be $h < 9.22 \times 10^{-3}$. Such Yukawa couplings can naturally satisfy the constraints from the requirement of light neutrino masses \eqref{cond1}.

Although the decay of $\psi \rightarrow \phi_2 L, \psi \rightarrow \chi \nu_R$ can also produce an abundance of dark matter (assumed to be the neutral component of $\phi_2$ here). However, since such scalar dark matter has large enough annihilation cross section to freeze out at a temperature much below the scale of leptogenesis $T = M_{\psi_1}$, the final dark matter relic abundance is not sensitive to such productions from the decay of $\psi$. We therefore calculate the dark matter relic abundance following the usual freeze out procedure outlined above. The different annihilation channels of the lighter neutral component of the doublet $\phi_2$ are shown as a function of its mass in the left panel of figure \ref{figdm1}. The typical annihilation cross section required to produce the correct relic abundance from the simplified formula \eqref{dm_relic} is shown as the horizontal line. This shows two different mass ranges: one below $M_W$, near the resonance $m_{DM} = m_h/2$ and another in the heavy mass regime, in agreement with earlier works on inert doublet dark matter mentioned earlier. It is worth noting that for the maximum value of Yukawa coupling $h$ consistent with successful leptogenesis for the chosen value of $\psi_1$ mass, the additional annihilation channel into neutrinos mediated by $\psi$ remains very much suppressed. In the right panel of \ref{figdm1}, we show the relic abundance by summing over all possible annihilation channels and for three different values of DM-Higgs couplings. We then chose the low mass regime, allowed from relic abundance criteria and show the allowed parameter space in $\lambda-m_{DM}$ plane in figure \ref{figdm2}. It can be seen that, the latest LUX bound on direct detection cross section and LHC bound on invisible Higgs decay significantly constrain the parameter space, leaving out only a tiny region near the resonance $m_{DM} = m_h/2$. It should be noted that the LUX 2016 bound appears as a band in the figure \ref{figdm2} after taking into the uncertainty in the Higgs nucleon couplings mentioned above.

Then we calculate the contribution to charged lepton flavour violating decay $\mu \rightarrow e \gamma$ from the charged component of the scalar doublet $\phi_2$. Taking the dark matter mass to be 62 GeV, allowed from all existing bounds (as seen from figure \ref{figdm2}), we first calculate the bound on charged scalar mass from EWPD. This together with LEP bound only leaves a small allowed region shown in figure \ref{figlfv}. We also show the exclusion line from MEG 2016 bound on $\text{BR}(\mu \rightarrow e \gamma)$. We then show the branching ratio as a function of charged scalar mass for two different choices of Yukawa coupling $h$ and the chosen value of $\psi_1$ mass, $M_{\psi_1} = 5$ TeV. It can be seen from figure \ref{figlfv} that for the maximum allowed value of Yukawa (from successful leptogenesis), the contribution to the LFV decay remains suppressed by more than three order of magnitudes compared to the latest MEG bound. However, for higher values of Yukawa, the corresponding branching ratio lies very close to the MEG bound and should be observed in near future. 

Finally, to have an overall idea about the allowed parameter space of the model from all phenomenological considerations discussed in this work, we intend to do a parameter scan incorporating the constraints from neutrino mass, leptogenesis, dark matter, LFV and EDM. The corresponding parameter space is huge which makes the complete scan very difficult. Instead of this complicated full scan, we therefore perform a partial scan in a subset of parameters while keeping the other parameters in a range that gives correct phenomenology. First, we keep the dark matter mass in the low mass regime shown in figure \ref{figdm2}. This also fixes the mass of the charged component of the scalar doublet $\phi_2$ to a narrow range from LEP constraints as seen from figure \ref{figlfv}. Keeping dark matter and $\phi^{\pm}_2$ mass in this range, we scan through other relevant parameters $h, M_{\psi}, \lambda_{\phi^-_2}$ from the requirement of leptogenesis, LFV and EDM. The resulting parameter space from this partial scan can satisfy constraints from neutrino mass if we tune other parameters involved in the neutrino mass formula \eqref{numassR} appropriately. In figure \ref{figlfv2}, we show the parameter space in terms of $(h, M_{\psi})$ that gives rise to successful leptogenesis with the heavy fermion decay freezing out before the sphaleron transition as well as agreement with the constraint from the bounds on charged lepton flavour violating decay $\mu \rightarrow e \gamma$. It can be seen from the figure \ref{figlfv2} that, these requirements put a strict lower bound on the heavy fermion mass $M_{\psi} \geq 4$ TeV. We then take into account the experimental constraint on electron EDM and show the allowed parameter space in terms of $(\lambda_{\phi^-_2}, M_{\psi})$ in figure \ref{figlfv3} which also satisfy the leptogenesis and LFV bounds.

It is interesting to note that, successful leptogenesis can occur in this scenario at TeV scale without any resonance enhancement like in the case of resonant leptogenesis \cite{resonant}. In usual vanilla leptogenesis, the same Yukawa coupling controls the lepton asymmetry, light neutrino mass as well as out-of-equilibrium condition of the heavy neutrino decay. To produce the correct lepton asymmetry and to satisfy the out-of-equilibrium criteria pushes the lightest right handed neutrino mass beyond $10^9$ GeV \cite{DIbound}. For TeV scale right handed neutrino, the out-of-equilibrium condition can be satisfied only for very small values of Yukawa couplings, of the order of $10^{-6}$, insufficient for producing the required asymmetry. This can be compensated however, by a mass degeneracy between two right handed neutrinos, giving rise to the scenario of resonant leptogenesis \cite{resonant}. In the present model, the most important out-of-equilibrium condition is the one that keeps the left and right sector away from equilibration, given in equation \eqref{outofeqb}. Since this depends on the fourth power of Yukawa couplings, compared to the second power of Yukawas in case of decay width in vanilla leptogenesis, one can satisfy the condition \eqref{outofeqb} with moderately large values of Yukawa couplings, which are also sufficient to produce the correct baryon asymmetry, as seen in our work by solving the explicit Boltzmann equations. Also, unlike in type I seesaw leptogenesis, here the constraints from light neutrino mass can be relaxed to some extent as the Yukawa coupling $h$ and heavy neutrino mass $M_{\psi}$ are not the only parameters that decide light neutrino masses, due to the involvement of other parameters in the one-loop neutrino mass diagram.

To summarise, we have studied a model which can simultaneously address the the problem of tiny neutrino mass, dark matter and baryon asymmetry with the new physics sector lying close to the TeV scale. Apart from being in agreement with all observed data, the model also offers the possibility of being verified or falsified at ongoing and near future experiments. While the LUX 2016 data allow a very tiny window of dark matter mass in the low mass regime, the EWPD constraints fixes the charged scalar mass to a small range at the same time. Although direct detection experiments may take a long time to rule out the resonance region completely, the LFV decay and electron EDM provide an alternate probe of that mass range. Near future observation of $\mu \rightarrow e \gamma$ or electron EDM will be able to constrain the currently allowed parameter space further. One interesting way to falsify this model is to look for $0\nu\beta\beta$ as it will rule out the pure Dirac nature of light neutrinos considered in this particular model.

\begin{acknowledgments}
DB would like to express a special thanks to the Munich Institute for Astro and Particle Physics (MIAPP) for hospitality and support while initial part of this work was completed.
\end{acknowledgments}

\appendix
\section{EDM Calculations}
\label{appen1}
The EDM of charged lepton at two-loop is given by 
\begin{equation}
 d_l = 2\Im \left[\frac{Y_l \lambda_{\phi^-_2} v h^2 }{256 \pi^4} (\sum^2_{i=1}(\sigma^i_R 
	- \sigma^i_L)  )\right] 
\end{equation}
where 
\begin{align}
	  \sigma^1_L &= m^2_l \int \prod^6_{i=1} dx_i \delta(1-\sum^6_{j=1}x_j)
	  \left[ \frac{U_1^{-3}}{F_1}\left(\frac{2}{F_1}\left((x_2 + x_3)(x_4 + x_5) + (x_2 + x_3 + x_4)x_6\right) \right.\right. \nonumber \\
	      &\left.\left.\left(x^2_2x_6(x_4 + x_5 +x_6) + x_2x_6(x_4x_6 + x_3(x_4 
		    + x_5 + x_6)) + ((x_1 + x_2 + x_3)x_4 + (x_3 + x_4)x_6)\right) \right.\right. \nonumber \\
	      &- \left. \left. \left(\frac{x_6((x_2+x_3)(x_4 + x_5) + x_6(x_2 + x_3 + x_4))}{(x_1 + x_2 + x_3)(x_4 + x_5) + x_6(x_1 + x_2 + x_3 + x_4 + x_5)}\right)\right)
	   + \frac{U_1^{-2}}{F^2_1}\left[x^2_2x_6(x_4 + x_5 + x_6) 
	        \right. \right. \nonumber \\
		 &+ x_2x_6(x_4x_6 + \left.\left. x_3(x_4 + x_5 x_6)) + ((x_1 + x_2 + x_3)x_4 
		  + (x_3 + x_4)x_6)(x_4x_6 + x_3(x_5 + x_4 + x_6)) \right. \right.\nonumber \\
		  &- \left. \left. \frac{F_1}{2}\frac{x_6}{(x_1 + x_2 + x_3 + x_6)(x_4 + x_5 + x_6) - x^2_6}\right] \right] \\
		   \sigma^1_R &= m^2_l \int \prod^6_{i=1} dx_i \delta(1-\sum^6_{j=1}x_j)
		   \left[ \frac{U_1^{-3}}{F_1} \left(2x_2x_4((x_2 + x_3)(x_4 + x_5) + (x_2 + x_3 + x_4)x_6)((x_1 + x_2 + x_3) \right. \right. \nonumber \\
			 & + (x_4 + x_5)+ \left. \left. x_6(1-x_6))\right) +  \frac{U_1^{-2}}{F_1^2}\left[x_2x_4((x_1 + x_2 + x_3)(x_4 + x_5) + x_6(1-x_6) )\right]
		   \right] 
\end{align}
\begin{align}
		   \sigma^2_L &= m^2_l \int \prod^6_{i=1} dx_i \delta(1-\sum^6_{j=1}x_j) \frac{U_2^{-3}}{F^2_3}\left(-(x_4 + 2x_5)x_6((x_1  + x_2)(x_3 - x_5) + (x_1 + x_2 
			   + x_3 - x_5)x_6 \right. \nonumber \\
			 &- \left. (x_4 + 2x_5)x_6(x_1(x_5-x_6) + x_5(x_2+x_6)) 
			 + (x_1(x_5 - x_6) + x_5(x_2 + x_6))((x_3 + x_4 + 2x_5)x_6 \right. \nonumber \\
			 & +\left. x_2(x_3 + x_4 + 
			     x_5 + x_6)) )-\frac{F_2}{2}\frac{x_6}{(x_1 + x_2 + x_6)(x_3 + x_4 + x_5 +x_6 
			   )-x^2_6}\right)
\end{align}
\begin{align}			   
		   \sigma^2_R &= m^2_l \int \prod^6_{i=1} dx_i \delta(1-\sum^6_{j=1}x_j)\frac{U_2^{-2}}{F^2_3}\left(((x_1 +x_2)(x_3 +x_4 + x_5) + (x_1 +x_2 + x_3 + x_4 + x_5)x_6)\right. \nonumber \\
		       &\left.(x_3x_6 + x_2(x_3-x_5 + x_6))\right) \\
		         U_1 &= \left(x_1+x_2+x_3\right)\left(x_4 + x_5\right) + x_6\left(x_1 + x_2+ x_3 + x_4+ x_5\right)
\end{align}
\begin{align}		         
	 U_2 &= (x_1 + x_2) (x_3 + x_4 + x_5) + (x_1 + x_2 + x_3 + x_4 + x_5) x_6 \\
	 F_1 &= m^2_lx_4 \left(\left(x_1 + x_2 x_3\right)x_4 + \left(x_2 + x_3 + x_4\right)x_6\right) + m^2_l\left(x_2 + x_3\right) \left(\left(x_2 + x_3\right) \left(x_4 + x_5\right) + \left(x_2 + x_3 + x_4\right) x_6\right) \nonumber \\
	 & + \left(\left( x_1 + x_2 + x_3 \right)\left(x_4 + x_5\right) + \left(x_1 + x_2 + x_3 + x_4 + x_5\right) x_6\right) \left(M_\psi^2 x_1 \right. \nonumber \\
	     & \left.+  m_{\phi^0_1}^2 x_4 
	     - m_l^2 \left(x_2 + x_3 + x_4 - x_5\right) + m_{\phi^0_2}^2 \left(x_2 + x_3 
	       + x_6\right)\right) \\
	 F_2 &= m_l^2 x_3 ((x_1 + x_2) x_3 + (x_2 + x_3) x_6) + ((x_1 + x_2) (x_3 + x_4 + x_5) + (x_1 + x_2 + x_3 + x_4 + x_5) x_6) \nonumber \\
	 &(M_\psi^2 x1 + m_{\phi^0_1}^2 x_4 - 
	         m_l^2 (x_2 + x_3 - x_5) + m_{\phi^-_2}^2 (x_2 + x_3 + x_6)) + 
	  m_l^2 x_2 (x_3 x_6 + x_2 (x_3 + x_4 + x_5 + x_6))
	  		   \end{align}

\bibliographystyle{JHEP}

\end{document}